\documentclass[aps,prd,twocolumn,showpacs,nofootinbib]{revtex4}
\usepackage{epsfig,graphicx,amssymb,bm}
\newcommand{\simgt}{\lower.5ex\hbox{$\; \buildrel > \over \sim \;$}}
\newcommand{\simlt}{\lower.5ex\hbox{$\; \buildrel < \over \sim \;$}}

\newcommand{\MNRAS}{Mon. Not. Roy. Astr. Soc.}

\newcommand{\skaco}[1]{\langle{#1}\rangle}

\begin{document}

\title{%
Can a galaxy redshift survey  measure dark energy clustering?
}%
\author{%
 Masahiro Takada
}%
\email{takada@astr.tohoku.ac.jp}
\affiliation{%
Astronomical Institute, Tohoku University, Sendai 980-8578, Japan
}%


\begin{abstract}
 A wide-field galaxy redshift survey allows one to probe galaxy
 clustering at largest spatial scales, which carries invaluable
 information on horizon-scale physics complementarily to the cosmic
 microwave background (CMB).  Assuming the planned survey consisting of
 $z\sim 1$ and $z\sim 3$ surveys with areas of $2000$ and 300
 deg$^2$, respectively, we study the prospects for probing dark
 energy clustering from the measured galaxy power spectrum, assuming the
 dynamical properties of dark energy are specified in terms of the
 equation of state and the effective sound speed $c_{\rm e}$ in the
 context of an adiabatic cold dark matter dominated model. The dark energy
 clustering adds a power to the galaxy power spectrum amplitude at
 spatial scales greater than the sound horizon, and the enhancement is
 sensitive to redshift evolution of the net dark energy density,
 i.e. the equation of state. We find that the galaxy survey, when
 combined with CMB expected from the Planck satellite mission, can
 distinguish dark energy clustering from a smooth dark energy model such
 as the quintessence model ($c_{\rm e}=1$), when $c_{\rm e}\simlt 0.04$
 (0.02) in the case of the constant equation of state $w_0=-0.9$
 ($-0.95$). An ultimate full-sky survey of $z\sim 1$ galaxies allows the
 detection when $c_{\rm e}\simlt 0.08$ (0.04) for $w_0=0.9$
 ($-0.95$). These forecasts show a compatible power with an all-sky CMB
 and galaxy cross-correlation that probes the integrated Sachs-Wolfe
 effect.  We also investigate a degeneracy between the dark energy
 clustering and the non-relativistic neutrinos implied from the neutrino
 oscillation experiments, because the two effects both induce a
 scale-dependent modification in the galaxy power spectrum shape at
 largest spatial scales accessible from the galaxy survey. It is shown
 that a wider redshift coverage can efficiently separate the two effects
 by utilizing the different redshift dependences, where dark energy
 clustering is apparent only at low redshifts $z\simlt 1$.
\end{abstract}
\maketitle

\section{Introduction}
Various cosmological probes such as supernovae in distant galaxies
\cite{Perlmutter99,Riess99}, the cosmic microwave background (CMB) sky
\cite{WMAP3}, and the galaxy redshift surveys
\cite{Tegmark04,2dF,Scranton,Eisenstein} have given strong evidence that
a dark energy component, such as the cosmological constant, constitutes
approximately $70\%$ of the total energy density of the universe, which
derives the accelerating cosmic expansion at low redshifts.  Because
there is no plausible theoretical explanation for its existence and
magnitude (e.g. see \cite{weinberg89,Ratra}), observational
exploration of the nature of dark energy is one of the most important
issues in modern cosmology as well as particle physics.

An observational dark energy task we should first explore to tackle this
fundamental problem would be to determine whether the accelerating
expansion is as a consequence of the cosmological constant.  Relaxing
this assumption leads to a generalized dark energy with dynamically
evolving energy density, which can be characterized by a time-dependent
equation of state $w(a)=p_{\rm de}/\rho_{\rm de}$ (the cosmological
constant corresponds to $w=-1$). The current level of accuracy in
constraining the constant $w$-parameter is $\sigma(w)\sim 0.1$ at
$1\sigma$ level with the best-fit value containing a model with $w=-1$
(e.g. see \cite{Eisenstein,SNLS,WMAP3}). Future prospects aimed at
pinning down the constraint on $w$ by a factor of 10 have been
extensively investigated to address the usefulness of various
cosmological experiments based on massive galaxy surveys such as
tomographic weak lensing experiment (e.g. \cite{Huterer,TJ}), the baryon
oscillation experiment (e.g.  \cite{SE,HuHaiman,taka05}) and the cluster
abundance experiment (e.g.  \cite{Laura}).

Another important consequence of a generalized dark energy is the
spatial perturbation, providing an independent clue to resolving the
nature of dark energy from the equation of state.  There are many
previous efforts made to study how an inclusion of the dark energy
perturbations modify the cold dark matter (CDM) structure formation
scenarios: theoretical predictions on the modifications in the CMB power
spectra and the galaxy power spectra
\cite{Caldwell,Chiba,Garriga,Dave,Ma,Hu02,Moroi,Huterer,Dedeo,Uzan} and the
observational exploration of the dark energy clustering signal from the
WMAP data at low multipoles \cite{WMAP3,Bean,Lewis,GordonHu}. The dark
energy clustering is relevant only for structure formation at low
redshifts $z\simlt 1$, where the net energy density is dominant to the
cosmic expansion.  In addition, a reasonable model including the dark
energy perturbation predicts that the dark energy can cluster together
with matter components at large spatial scales (inevitably at
super-horizon scales), whilst the dark energy is smooth at small scales
so that the additional component does not largely change the small-scale
structure formations such as galaxy formation.
For these reasons, the dark energy clustering effect on the CMB
observables is likely to appear only via the integrated Sachs-Wolfe
effect (ISW) at low multipoles, where the Sachs-Wolfe effect generated
at the recombination epoch is significant contamination to separate
the ISW effect from the measured power spectrum.

The transition scale to divide the dark energy clustering and smooth
regimes can be usefully modeled by the effective sound speed of dark
energy \cite{Hu98,Hu02}. In this model, smoothness of dark energy can be
tested by searching for the signature of the sound speed from
cosmological observables. Hu and Scranton \cite{HuScranton} carefully
investigated a prospect of how the ISW effect measured via the angular
cross-correlation between CMB and galaxy distribution can be used to
probe the dark energy clustering, assuming an all-sky, deep multi-color
imaging galaxy survey out to $z\sim 2$.

A galaxy redshift survey offers an alternative means for probing the
dark energy clustering, through the measured statistical properties of
three-dimensional gravitational clustering at largest scales.  Compared
to the angular correlations, from a cosmological point of view, the
redshift survey carries more information on the underlying mass
distribution due to a gain of the modes along the line-of-sight or
redshift direction.  Therefore, the purpose of this paper is to, for the
first time, investigate the ability of a galaxy redshift survey for
testing the smoothness of dark energy from the measured galaxy power
spectrum.  In fact, there are several future plans for high-redshift
galaxy surveys that are already being constructed or seriously under
consideration: the Fiber Multiple Object Spectrograph (FMOS) on Subaru
telescope \cite{fmos}, its significantly expanded version, WFMOS
\cite{wfmos}, the Hobby--Ebery Telescope Dark Energy eXperiment (HETDEX)
\cite{hetdex}, and the Cosmic Inflation Probe (CIP) mission \cite{cip}.
These surveys probe galaxies at higher redshifts $z\simgt 0.5$ than the
existing surveys such as SDSS and 2dF surveys. Such a high-redshift
survey has several advantages over the lower redshift ones. First, given
a fixed solid angle, the comoving volume in which we can observe
galaxies is larger at higher redshifts than in the local universe,
thereby reducing the sample variance. This would make it more
straightforward to obtain a well-behaved survey geometry that can help
measure largest-scale perturbations. Second, density perturbations at
smaller spatial scales are still in the linear regime or only in the
weakly non-linear regime at higher redshift, which gives us more
leverages on measuring the shape of the linear power spectrum to break
the parameter degeneracies. In this paper, we will consider the survey
design close to the proposed WFMOS survey, which consists two types of
surveys different in redshift coverage and survey area: $0.5\le z\le
1.3$ with 2000 degree$^2$ and $2.5\le z\le 3.5$ with $300$ degree$^2$,
respectively. The redshift coverage of WFMOS is suitable to probe the
dynamical dark energy whose effects are apparent only at low redshifts
$z\simlt 1$, as we will show below.

The structure of this paper is as follows. In Sec.~\ref{prel} we start
with writing down background cosmological equations, the Hubble
expansion and the angular diameter distance, in terms of cosmological
parameters. In Sec.~\ref{de}, we define the effective sound speed
parameter to model dynamical properties of dark energy clustering, and
review how the dark energy clustering leads to a scale-dependent
modification in the linear power spectrum shape assuming the adiabatic
initial condition. Sec.~\ref{gps} defines the galaxy power spectrum in
terms of the primordial power spectrum, the transfer functions and the
scale-dependent growth rate of mass clustering. In Sec.~\ref{method}, we
first define survey parameters intended to resemble a future survey
being planned, and describe a methodology to model the galaxy power
spectrum observed from a redshift survey that includes the
two-dimensional nature in the line-of-sight and transverse directions
due to the cosmological and redshift distortion effects.  We then
present the Fisher information matrix formalism that is used to estimate
the projected uncertainties in the cosmological parameter determination
provided the measured galaxy power spectrum. In Sec.~\ref{results} we
show the prospects of the future survey for probing the dark energy
clustering.  In addition, we carefully study how a degeneracy between
the dark energy clustering and massive neutrinos can be lifted by
utilizing the redshift information of galaxy clustering.  Finally, we
present conclusion and some discussion in Sec.~\ref{conc}.

\section{Preliminaries: cosmology}
\label{prel}

Throughout this paper, we work in the context of spatially flat CDM
models for structure formation (e.g. see \cite{dodelson} and
\cite{PeacockText}). According to the Einstein general relativity, the
expansion history of the universe is given by the scale factor $a(t)$,
which is related to redshift via $1+z=1/a$ (we use $a(t_0)=1$ today for
our convention). The cosmic expansion during the matter dominated epoch
is determined by density contributions from non-relativistic matter
$\Omega_{\rm m}$ and dark energy $\Omega_{\rm de}$ at present, in units
of the critical density $3H_0^2/8\pi G$, where $H_0=100~ h~ $km
s$^{-1}$Mpc$^{-1}$ is the Hubble constant. The Hubble expansion rate is
given by
\begin{equation}
H^2(a)=H_0^2\left[\Omega_{\rm m}a^{-3}+\Omega_{\rm
	     de}a^{-3(1+w_0)}\right], 
\end{equation}
where we have assumed the constant equation of state of dark energy,
\begin{equation}
w_0\equiv \frac{p_{\rm
de}(a)}{\rho_{\rm de}(a)}=-
\frac{1}{3}\frac{d\ln\rho_{\rm de}}{d\ln a}-1={\rm const.}
\end{equation}
Note that $w_0=-1$ corresponds to the cosmological constant. In this
paper we restrict ourselves to a dark energy model with $w_0\ge -1$ for
simplicity. 

The comoving angular diameter distance is expressed in terms of the
Hubble parameter as  
\begin{equation}
D_A(a)=\int^a_1\!\frac{da}{H^2(a')a^{\prime 2}}, 
\end{equation}
giving the distance-redshift relation via $1+z=1/a$.

\section{Dark Energy Clustering for Adiabatic Initial Condition}
\label{de}

To model dark energy clustering, we employ a phenomenological model
developed in \cite{Hu02}.  In this model, the stress perturbation of
dark energy, which governs properties of the dark energy clustering, is
usefully specified by the dark energy equation of state and the
effective sound speed $c_{\rm e}$, where the latter is needed to model
the non-adiabatic stress perturbation (also see \cite{Chiba}). The
effective sound speed of a generalized dark energy is defined as
\begin{equation}
c_{\rm e}^2\equiv \left.\frac{\delta p_{\rm de}}{\delta \rho_{\rm
		     de}}\right|_{\rm rest},
\label{eqn:ce}
\end{equation}
in a ``rest frame'' coordinate system where the momentum density of the
dark energy vanishes. For a more general coordinate system such as
Newtonian gauge, the pressure perturbation of dark energy, which enters
into the r.h.s. of the momentum conservation equation, can be
expressed as
\begin{eqnarray}
\delta p_{\rm de}
&=&w_0 \delta\rho_{\rm de}+\bar{\rho}_{\rm de}
(c_{\rm e}^2-w_0)\left(\delta_{\rm de}+3\frac{\dot{a}}{a}\frac{u_{\rm de}}{k}\right)
\nonumber \\
&=& c_{\rm e}^2\delta\rho_{\rm de}+
3\frac{\dot{a}}{a}(c_{\rm e}^2-w_0)
\frac{u_{\rm de}}{k}\bar{\rho}_{\rm de}, 
\label{eqn:pde}
\end{eqnarray}
where the first and second terms in the first line on the r.h.s.  denote
the adiabatic and non-adiabatic pressure perturbations, respectively,
$u_{\rm de}$ denotes the peculiar velocity of dark energy, and we have
assumed $w_0={\rm constant}$ in time.  Note that setting $u_{\rm de}=0$
in Eq.~(\ref{eqn:pde}) (corresponding to the rest frame of dark energy
perturbations) reduces the pressure perturbation to the form given by
Eq.~(\ref{eqn:ce}). From Eq.~(\ref{eqn:pde}), one can find that presence
of the effective sound speed leads the pressure perturbation to act as a
restoring force against the gravitational instability just like the
Jeans instability of baryon perturbation (e.g., see Sec.~16 in
\cite{Peebles80}); in the overdensity ($\delta_{\rm de}>0$) and
underdensity ($\delta_{\rm de}<0$) regions, the pressure perturbation
prevents further collapse and expansion, respectively, if the wavelength
of perturbation is smaller than the sound horizon (see below).
Conversely, for a model that has only the adiabatic pressure
perturbation $(\delta p_{\rm de}=w_0\delta\rho_{\rm de})$, the pressure
perturbation and the gravitational force in the momentum conservation
equation become $kw_0\delta_{\rm de}+(1+w_0)k\Psi$ ($\Psi$ is the
gravitational potential), namely, the two terms have same sign because
$w_0<0$ and $\Psi\propto -\delta_{\rm de}$ via the Poisson equation:
therefore, the pressure perturbation accelerates the gravitational
instability on all scales. Note that, throughout this paper, we assume
the constant sound speed with $c_{\rm e}\ge 0$ and ignore the trace-free
stress perturbation of dark energy for simplicity.

We expect that this modeling to treat the sound speed as a free
parameter can cover a broader range of dark energy models rather than
working with a specific model of the Lagrangian of dark energy sector.
For a scalar field dark energy model, the sound speed can be exactly
computed in linear theory in terms of the kinetic energy as a function
of the field \cite{Garriga,GordonHu}. For the case of a canonical
kinetic energy term such as the quintessence
\cite{Ratra,Frieman95,Caldwell,Chiba}, the sound speed $c_{\rm e}=1$. For a
more general modification of the kinetic term such as a phantom energy
\cite{Caldwell02} and a k-essence field \cite{chiba,kessence}, a general
value of $c_{\rm e}$ including $c_{\rm e}\ll 1$ is allowed.  

The dark energy sound speed sets a characteristic length scale in
gravitational clustering:
\begin{equation}
\lambda_{\rm de, fs}(a)=\int_{0}^a\!da~ \frac{c_{\rm e}}{a^2H(a)}.
\label{eqn:defs}
\end{equation}
We shall often call this scale the comoving sound horizon of dark energy
at epoch $a$. One might define the corresponding wavenumber as $k_{\rm
de,fs}=2\pi/\lambda_{\rm de,fs}$. For the scalar-field dark energy model
with canonical kinetic energy ($c_{\rm e}=1$), the sound horizon is
comparable with the particle horizon scale. In this paper, we simply
refer to a model with $c_e=1$ as a smooth dark energy model, because a
galaxy survey we consider can probe only the galaxy clustering at scales
inside the present-day horizon scale and there is no unique definition
of smoothness of dark energy on super-horizon scales where the energy
density varies depending on the freedom of time-slicing in general
relativity.

On large scales $\lambda>\lambda_{\rm de,fs}$, the dark energy
perturbation can grow together with non-relativistic matter
components. In this case, the gravitational potential $\Phi$ in
Newtonian gauge\footnote{More precisely, $\Phi$ is the curvature
perturbation, but related to the potential perturbation via $\Phi=-\Psi$
when the trace-free stress perturbation is negligible, which is valid at
redshifts of our interest.}  evolves in a flat universe (e.g., see
\cite{HE99prd}) as
\begin{equation}
\Phi(\bm{k},a)=\phi_c(a){\cal R}_i(\bm{k})
\end{equation}
where 
\begin{equation}
\phi_c(a)\equiv \left(1-\frac{H(a)}{a}\int^a_0\!da'\frac{1}{H(a')}\right)
\label{eqn:phi_c}
\end{equation}
and ${\cal R}_i(\bm{k})$ denotes the primordial curvature perturbation
of wavenumber $\bm{k}$. In the matter dominated regime, $H(a)\propto
a^{-3/2}$, leading to $\phi_c\rightarrow 3/5$.  Here we have employed
the adiabatic initial condition for dark energy perturbations; the
initial conditions of all the components are set by the primordial
curvature perturbation.  It is worth noting that, for a cosmological
constant model ($w_0=-1$), Eq.~(\ref{eqn:phi_c}) can be rewritten as
\begin{eqnarray}
1-\frac{H(a)}{a}\int_0^a\!\frac{da'}{H(a')}&=&1-\frac{H(a)}{a}
\left[\left.\frac{a'}{H(a')}\right|^a_0\right.\nonumber\\
&&\hspace{0em}\left. +
\int^a_0\!da'\frac{ a'}{2H^3(a')}\frac{dH^2(a')}{da'}
\right]\nonumber\\
&\propto &\frac{H(a)}{a}\int^a_0\!da'\frac{1}{H^3(a')a^{\prime 3}},
\label{eqn:phi_c2}
\end{eqnarray}
where we have used the partial integral in the first line on the r.h.s.,
and $dH^2(a)/da=-3 H_0^2\Omega_{\rm m0} a^{-4}$ in the second line.  The
final expression is equivalent to the well-known formula for the growth
rate of mass clustering, for example, given by Eq.~(10.12) in
\cite{Peebles80} (also see \cite{Heath}).

On the other hand, on small spatial scales $\lambda<\lambda_{\rm de,
fs}$, the pressure perturbation of dark energy prevents clustering of
the dark energy; the dark energy is thus considered as a smooth
component in this limit.  The redshift dependence of the gravitational
perturbation is given by $\Phi\propto g(a)$ via $1+z=1/a$, and the
growth rate $g(a)$ can be computed (e.g. see \cite{TJ}) by solving the
following differential equation with the initial conditions $g(a_{\rm
md})=1$ and $dg/da(a_{\rm md})=0$ in the deeply matter dominated regime
$a_{\rm md}$ (e.g., $a_{\rm md}=10^{-3}$):
\begin{eqnarray}
&&\hspace{-5em}2\frac{d^2g(a)}{d\ln a^2}+\left[5-3w(a)\Omega_{\rm de}(a)\right]
\frac{dg(a)}{d\ln a}\nonumber\\
&&\hspace{2em}+3\left[1-w(a)\right]\Omega_{\rm de}(a)g(a)=0,
\label{eqn:g}
\end{eqnarray}
where the equation of state is generalized so that it is allowed to have
a time dependence $w(a)$, and $\Omega_{\rm de}(a)$ is the dark energy
density parameter at epoch $a$, defined as $\Omega_{\rm de}(a)\equiv
8\pi G\rho_{\rm de}(a)/3H^2(a)$. For a cosmological constant model, the
solution of Eq.~(\ref{eqn:g}) can be given by the integral form of
Eq.~(\ref{eqn:phi_c}).  Note that in the presence of the
non-relativistic neutrinos the growth rate is further modified at scales
below the neutrino free streaming scale (see \cite{TK}), as we shall
discuss below.

Hence, introducing the effective sound speed of the dark energy enforces
that the stress perturbation leads the dark energy perturbation to be
gravitationally stable on small scales so that small-scale structure
formations are not largely modified by the additional component compared
to the concordance $\Lambda$CDM predictions.

\section{Shape of Linear Galaxy Power Spectrum}
\label{gps}

To model the linear power spectrum including the dark energy
perturbation, we use the recipe in \cite{Hu02} to compute the transfer
function of the potential perturbation $\Phi(k,z)$ assuming the
adiabatic initial condition.  Given the primordial power spectrum, the
power spectrum of the potential perturbation in the matter dominated
regime is given by
\begin{eqnarray}
\hspace{-1em}\Delta_\Phi^2(k,z)&=&  T_{\rm de}^2(k,z)
\frac{9}{25}\delta_{{\cal R}}^2 \nonumber\\
&&\hspace{-5em}\times \left(\frac{D_{cb\nu}(k,z)}{a}\right)^2T^2(k)\left(
\frac{k}{k_0}
\right)^{-1+n_s+\frac{1}{2}\alpha_s\ln(k/k_0)}, 
\label{eqn:Pphi}
\end{eqnarray}
where the primordial power spectrum, $P_{{\cal R}}(k)\equiv \skaco{{\cal
R}_i^2}$, is specified in terms of the primordial curvature perturbation
$\delta_{\cal R}$, the spectral tilt $n_s$ and the running index
$\alpha_s$, and $T_{\rm de}(k,z)$ is the fitting formula given in
\cite{Hu02} to describe the dark energy clustering contribution as
explained below.  Note that the primordial power spectrum shape is
defined at $k_0=0.05$~Mpc$^{-1}$. In Eq.~(\ref{eqn:Pphi}), we have
imposed that the super-horizon potential perturbation is related to the
primordial curvature perturbation as $\Phi(z_{\rm md})=(3/5){\cal R}_i$
in the matter dominated regime.  The growth rate for the total matter
(CDM, baryon plus non-relativistic neutrinos) perturbations are computed
using the recipe developed in \cite{HE99} and normalized as
$D_{cb\nu}(k,z)\rightarrow a g(z)$ at $k\rightarrow 0$.  It is also
noted that, throughout this paper, we employ the transfer function of
matter perturbations, $T(k)$, with baryon oscillations {\it smoothed
out} for simplicity (see Sec.~IV.B in \cite{TK} for the related
discussion).

According to the physical processes described in Sec.~\ref{de}, the fitting
function $T_{\rm de}(k,z)$ in Eq.~(\ref{eqn:Pphi}) is given in terms of
the growth rates at scales smaller and larger than the dark energy sound
horizon as
\begin{equation}
T_{\rm de}(k,z)=\frac{1+q^2}{3g(z)/[5\phi_c(z)]+q^2},
\label{eqn:tde}
\end{equation}
where the variable $q$ is defined as
\begin{equation}
q\equiv \frac{k}{2\pi}\sqrt{\lambda_{\rm de,fs}(z)\lambda_{\rm de, fs}(z_{\rm de})},
\end{equation}
with the redshift $z_{\rm de}$ being given by
\begin{equation}
\frac{\rho_{\rm de}(z_{\rm de})}{\rho_{\rm m}(z_{\rm de})}= \frac{1}{\pi},\hspace{2em}
1+z_{\rm de}=\left(\pi\frac{\Omega_{\rm de}}{\Omega_{\rm
	      m}}\right)^{-\frac{1}{3w_0}}. 
\end{equation}
Note that $T_{\rm de}\rightarrow 5\phi_c(z)/[3g(z)]$ or $1$ as
$k\rightarrow 0$ or $\infty$, respectively, which reproduce the two
asymptotic regimes discussed in Sec.~\ref{de}. For the cosmological
constant model ($w_0=-1$), $T_{\rm de}(k,z)=1$ because
$3g(z)/[5\phi_c]=1$ as explained around Eqs.~(\ref{eqn:phi_c2}) and
(\ref{eqn:g}). 

Galaxies are biased tracers of the underlying gravitational
field. Hence, the linear power spectrum of galaxy distribution is
related to the potential power spectrum via the Poisson equation as
\begin{equation}
\frac{k^3 P_g(k,z)}{2\pi^2}=b_1^2\left(\frac{2k^2a}{3H_0^2\Omega_{\rm
				  m}}\right)^2
\Delta_\Phi^2(k,z)
\label{eqn:pg}
\end{equation}
where $b_1$ is a scale-independent, linear bias parameter we assume.

As carefully investigated in \cite{Hu02} (see Appendix A of that paper),
the fitting formula (\ref{eqn:tde}) holds with batter than $\sim 10\%$
accuracy compared to the results obtained by directly solving the
multi-fluid Boltzmann equations, for a range of the dark energy
parameters we are interested in. In the presence of non-relativistic
neutrinos, however, it remains unclear how our treatment of
Eq.~(\ref{eqn:Pphi}) holds accurate.  Hence, an interesting issue is to
develop an accurate transfer function for the total matter perturbations
including the dark energy as well as neutrino perturbations. This will
be our future study and presented elsewhere.  For our purpose, which is
to {\em estimate} the ability of future surveys to probe dark energy
clustering, our treatment of the galaxy power spectrum is accurate
enough. This is partly justified by the fact that the following results
are not largely changed with and without non-relativistic neutrinos, as
will be explicitly shown below.

\begin{figure}
  \begin{center}
    \leavevmode\epsfxsize=8.5cm \epsfbox{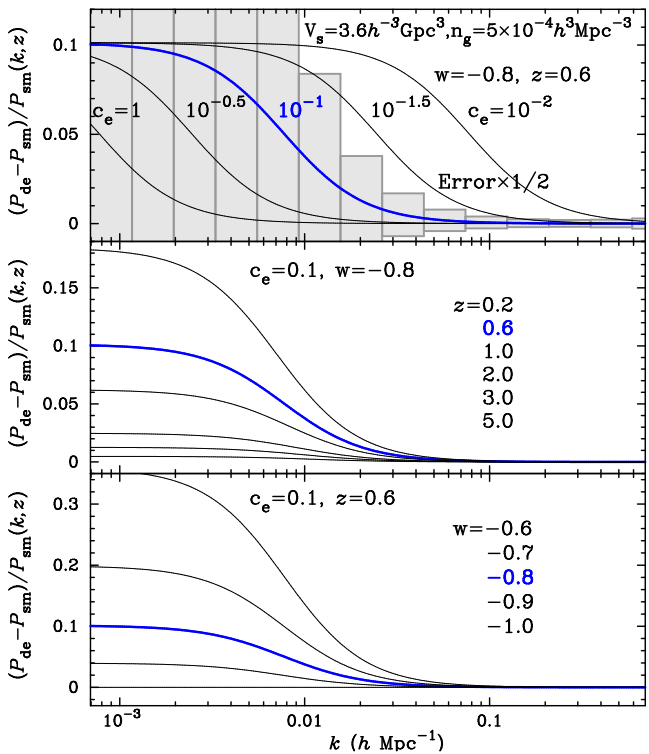}
  \end{center}
\vspace{-2em}
\caption{ \label{fig:pk} {\em Upper panel}: Shown is how the dark energy
 clustering amplifies the power of linear power spectrum $P(k)$ at
 $z=0.6$ on spatial scales greater than the dark energy sound horizon,
 relative to the smooth dark energy model ($P_{\rm sm}$).  Note that
 non-relativistic neutrinos are ignored in this plot.  The dark energy
 equation of state is taken to $w_0=-0.8$, and the sound speed is varied
 to show how the scale-dependent modification in $P(k)$ changes with
 $c_{\rm e}$: with decreasing $c_{\rm e}$, the transition in $P(k)$
 amplitude appears at smaller length scales, i.e. larger $k$. Shaded
 boxes around the curve of $c_e=0.1$ show the expected 1-$\sigma$
 measurement errors for spherically averaged $P(k)$ at each $k$ bin for
 a survey covering 2000 deg$^2$ and $0.5<z<1.3$ (see text for the
 details), but note that we have multiplied the error bars by a factor
 1/2 for illustrative purpose. Comparing the solid curves with the
 error bars leads to a naive expectation that the dark energy clustering
 can be detected from this type of galaxy survey only when the sound
 speed is smaller than $c_{\rm e}\sim 0.1$. {\em Middle panel}: Redshift
 dependence of the power spectrum shape modification due to the dark
 energy clustering, where $c_{\rm e}=0.1$ and $w_0=-0.8$ are fixed.
 {\em Lower panel}: As the fiducial value of $w_0$ is away from
 $w_0=-1$, the amplification in $P(k) $ at small $k$ is more enhanced,
 for a fixed $c_{\rm e}=0.1$.  In all the panels, the bold solid curve
 shows the same result for $c_e=0.1$, $w_0=-0.8$ and $z=0.6$. }
\end{figure}
The top panel of Fig.~\ref{fig:pk} illustrates how the dark energy
clustering induces a scale-dependent modification in the shape of linear
power spectrum, $P(k)$, at $z=0.6$, compared to the smooth dark energy
model.  The dark energy equation of state is fixed to $w_0=-0.8$, and
the fiducial value of the sound speed $c_{\rm e}$ is varied from $c_{\rm
e}=1$ to $c_{\rm e}=10^{-2}$.  Note that $c_{\rm e}=1$ roughly
corresponds to the smooth dark energy model such as the quintessence
model, since the sound horizon is comparable with the Hubble horizon.  It
is clearly seen that, on spatial scales larger than the dark energy
sound horizon given by Eq.~(\ref{eqn:defs}), the dark energy can cluster
together with dark matter and thus enhances the power spectrum amplitude
compared to the small-scale amplitude that matches the smooth model
prediction $(P/P_{\rm sm}=1)$. The transition in $P(k)$ amplitude
appears from larger $k$ with decreasing the sound speed $c_{e}$, because
the dark energy sound horizon gets shorter.  It is worth noting that,
since we have normalized the primordial power spectrum (see
Sec.~\ref{gps}), all the power spectra match at $k\ll k_{\rm de,{\rm
fs}}$ or $k\gg k_{\rm de, fs}$, showing that the effect of dark energy
perturbation becomes independent of $k$. Hence, one can measure $c_{\rm
e}$ only if the characteristic transition pattern in the power spectrum
shape is accurately measured from observations. In other words, the
power spectrum amplitude is not useful to constrain $c_{\rm e}$.  

The shaded boxes around the curve of $c_{\rm e}=0.1$ represent the
$1$-$\sigma$ measurement errors on $P(k)$, reduced by a factor $2$ for
illustrative purpose, for the fiducial galaxy survey with redshift range
of $0.5\le z\le 1.3$ and survey area of $2000$ deg$^2$ corresponding to
the comoving volume of $3.6h^{-3}$Gpc$^3$. Note that the errors are for
the power spectrum spherically averaged over angle.  It is apparent that
the future survey of our interest allows a precise measurement of the
linear power spectrum shape, achieving a few $\%$ accuracies in each
$k$-bin at $k\simgt 0.05h~$Mpc$^{-1}$, while the errors are dominated by
the sample variance at smaller $k$.  A quick look at this plot leads to
a naive expectation
that, only if the sound speed is sufficiently small such as $c_{\rm
e}\simlt 0.1$, one could measure the transition pattern in $P(k)$ from
the galaxy survey.  In the following, we shall carefully study how well
a future galaxy survey can probe the dark energy clustering fully taking
into account degeneracies between cosmological parameters.

In the middle panel we show that the dark energy perturbation more
amplifies the power spectrum amplitude at the large spatial scales as
one goes to a lower redshift, because the net dark energy density is
more prominent at lower redshifts. Observationally this means that a
galaxy survey of lower redshift $z\simlt 1$ is more suitable for
exploiting the dark energy perturbations than high-z ones.

Finally, the lower panel of Fig.~\ref{fig:pk} shows how the effect of
dark energy perturbation varies with the fiducial value of the equation
of state $w_0$.  With increasing $w_0$ from $w_0=-1$, the dark energy
perturbation more enhances the amplitude of $P(k)$ at the large scales,
because the dark energy density keeps prominent at higher redshift.

\section{Methodology}
\label{method}

\subsection{Survey Parameters}
\label{survey}

\begin{table*}
\begin{center}
\begin{tabular}{lccccccc}
\hline\hline & & $k_{\rm max} $ & $\Omega_{\rm survey}$ & $V_{\rm s}$ &
$\bar{n}_{\rm g}$ & \hspace{1em}Bias\hspace{1em} 
\\
Survey & $z_{\rm center}$ &($h$Mpc$^{-1}$) & (deg$^2$) &
($h^{-3}$Gpc$^3$) 
& ($10^{-3}~ h^3$Mpc$^{-3}$)& 
\\
\hline 
Gz1 ($0.5<z<1.3$) & $0.6$ & $0.15 $& 2000 & 0.57 & 0.5 & 1.25\\
    & $0.8$ & $0.17$ & 2000 & 0.81& 0.5 & 1.40\\
    & $1.0$ & $0.19$ & 2000 & 1.0 & 0.5 & 1.55\\
    & $1.2$ & $0.21$ & 2000 & 1.2 & 0.5 & 1.7\\
Gz3 ($2.5<z<3.5$) & $3.0$ & $0.53$& 300 & 1.2 & 1.0 & 3.3\\
\hline \hline
\end{tabular}
\end{center}
\caption{\label{tab:survey} Galaxy survey specifications that we assume
in this paper.  We consider two types of galaxy surveys, named as
``Gz1'' and ``Gz3'', intended to resemble the WFMOS fiducial survey
\cite{wfmos}. The former sees galaxies at $z\sim 1$ with a fixed sky
coverage of 2000 deg$^2$, while the latter probes $z\sim 3$ galaxies
with 300 deg$^2$.  $V_{\rm s}$ and $\bar{n}_{\rm g}$ are the comoving
survey volume and the comoving number density of sampled galaxies for
each redshift slice, respectively.  Note that $V_{\rm s}$ is computed
for our fiducial cosmological model with $\Omega_{\rm m}=0.27$,
$\Omega_{\rm de}=0.73$ and $w_0=-0.8$.  $z_{\rm center}$ denotes the
centering redshift of each redshift slice, and $k_{\rm max}$ is the
maximum wavenumber below which information in the linear power spectrum
can be extracted. (We do not use any information above $k_{\rm max}$ in
the Fisher information matrix analysis.) ``Bias'' denotes the assumed
linear bias parameters of sampled galaxies.}
\end{table*}

To derive a realistic parameter forecast, we employ the galaxy survey
parameters that are chosen to resemble the future surveys that are under
serious consideration.  As shown below (see Eq.~[\ref{eqn:pkerror}]),
the statistical error of the galaxy power spectrum measurement is
limited by the survey volume, $V_{\rm s}$, as well as the mean number
density of galaxies, $\bar{n}_g$.  There are two advantages for a higher
redshift survey over the current surveys probing the universe at
$z\simlt 0.3$.  First, given a fixed solid angle, the comoving volume in
which we can observe galaxies is larger at higher redshifts than in the
local universe, thereby reducing the sample variance.  In addition, it
would be relatively straightforward to obtain a well-behaved survey
geometry, e.g., a cubic geometry that would be helpful to probe a
largest-scale galaxy clustering as well as handle the systematics under
control.  Second, density fluctuations at smaller spatial scales are
still in the linear regime or only in the weakly non-linear regime at
higher redshift, which gives us more leverages on measuring the shape of
the linear power spectrum.

We employ the survey parameters that match the fiducial survey design of
WFMOS \cite{wfmos}, consisting of two types of redshift surveys
different in redshift coverage and survey area:
\begin{itemize}
\item Gz1: $0.5<z<1.3$ and $\Omega_{\rm s}=2000~$deg$^2$
\item Gz3: $2.5<z<3.5$ and $\Omega_{\rm s}=300~$deg$^2$
\end{itemize}
where the names Gz1 and Gz3 stand for the ``Ground''-based galaxy surveys
probing the universe at $z\sim 1$ and $z\sim 3$, respectively.

Because we have limited knowledge of how galaxies have formed within the
CDM hierarchical clustering scenario, it is of critical importance to
figure out an optimal survey strategy of which type of galaxies
(emission lines) are targeted to achieve the desired scientific goals,
given the spectrograph specifications (sensitivity, the number of
fibers, wavelength coverage, etc.) and the number of nights
allocated. It was shown in \cite{SE} (also see \cite{Blake}) that a
survey having $\bar{n}_{\rm g} P_{\rm g}\simgt 3$ over the range of
wavenumbers considered is close to an optimal design from a
cosmological, not practical, point of view, and the hypothetical WFMOS
survey parameters were defined in this regard.  In this paper, we simply
adopt the survey parameters used in \cite{SE}, which are summarized in
Table~\ref{tab:survey}.

\subsection{A Galaxy Power Spectrum in Redshift Space}
\label{ps}
We employ the method developed in \cite{SE} (also see \cite{TK}) to
model a galaxy power spectrum observable from a redshift survey.  The
power spectrum measures how clustering strength of galaxy distribution
varies as a function of 3-dimensional wavenumber, $k$ (or the inverse of
3-dimensional length scale).  However, we cannot know a true position of
a galaxy in real space from the observables, angular position and
redshift. Rather, we have to assume a fiducial cosmology to convert the
observed angular position and redshift of galaxies into positions in
3-dimensional space. Since this fiducial cosmology generally differs
from the true cosmology, we could introduce distortion in the inferred
distribution of galaxies. This cosmological distortion effect is the
so-called Alcock-Paczynski (AP) effect \cite{AP}.  Taking into account
the effects of the cosmological distortion, linear redshift distortion
\cite{Kaiser} and galaxy bias, the galaxy power spectrum in the linear
regime can be expressed in terms of the real-space galaxy power spectrum
$P_g(k,z)$ (see Eq.~[\ref{eqn:pg}]) as
\begin{eqnarray}
P_s(k_{{\rm fid}\perp},k_{{\rm fid}\parallel})
&=&\frac{D_A(z)^2_{\rm fid}H(z)}{D_A(z)^2H(z)_{{\rm fid}}}
\left[1+\beta \frac{k_\parallel^2}{k_\perp^2+k_\parallel^2}\right]^2\nonumber\\
&&\times P_g(k,z),
\label{eqn:ps}
\end{eqnarray}
with
\begin{equation}
k_\perp=\frac{D_A(z)_{{\rm fid}}}{D_A(z)} k_{{\rm fid}\perp}, 
\hspace{2em}k_\parallel=\frac{H(z)}{H_{\rm fid}(z)}k_{{\rm fid}\parallel},
\end{equation}
where
$k=(k_\perp^2+k_\parallel^2)^{1/2}$,  $k_\parallel$ and $k_\perp$
denote the components of the wavenumber parallel and perpendicular to
the line-of-sight direction, respectively, and $\beta$ denotes the
linear redshift space distortion taking into account the
non-relativistic neutrino effect (see \cite{TK}), defined as
$\beta=-(1/b_1)d\ln D_{\rm cb\nu}(k,z)/d\ln (1+z)$.  $D_A(z)$ and $H(z)$
are the comoving angular diameter distance and Hubble parameter,
respectively, and the quantities with subscript `fid' denote the
quantities in the fiducial cosmological model. Note that $P_g(k,z)$
depends on the scale-independent, linear bias parameter $b_1$ as
$P_g\propto b_1^2$ (see Eq.~[\ref{eqn:pg}]).

The redshift and cosmological distortion effects are extremely powerful
for constraining cosmological parameters. Because the structure formation
scenario predicts that the power spectrum $P(k)$ has characteristic features
such as the broad peak from the matter-radiation equality,
scale-dependent suppression of power due to baryons and non-relativistic
neutrinos, the tilt and running of the primordial power spectrum, and
the baryonic acoustic oscillations, the distortion effects can be
precisely measured with future galaxy surveys from the distorted
features and help break degeneracies between parameter determination
quite efficiently. There are notably two examples for the
prospects. First, the baryon oscillation peaks in $P(k)$ can be used as
a standard ruler, allowing precise measurements of $H(z)$ and $D_A(z)$
to constrain the equation of state of dark energy
\cite{SE,HuHaiman,Blake,taka05}. Second, measuring the redshift
distortion helps break strong degeneracy between the power spectrum
amplitude and the galaxy bias, allowing a precise determination of the
neutrino mass \cite{TK}.

\subsection{Fisher Matrix Analysis}

In order to investigate how well one can constrain the cosmological
parameters for a given redshift survey, one needs to specify measurement
uncertainties of the galaxy power spectrum.  When non-linearity is weak,
it is reasonable to assume that observed density perturbations obey
Gaussian statistics. In this case, there are two sources of statistical
errors on a power spectrum measurement: the sampling variance (due to
the limited number of independent wavenumbers sampled from a finite
survey volume) and the shot noise (due to the imperfect sampling of
fluctuations by the finite number of galaxies). To be more specific, the
statistical error on measurement of $P(k)$ at a given wavenumber bin
$k_i$ is given in \cite{FKP} by
\begin{equation}
\left[\frac{\Delta P_s(k_i)}{P_s(k_i)}\right]^2=
\frac{2(2\pi)^2 }{V_{\rm s}k^2\Delta k\Delta \mu}
\left[1+\frac1{\bar{n}_gP_s(k_i)}\right]^2,
\label{eqn:pkerror}
\end{equation}
where $\bar{n}_g$ is the mean number density of galaxies, $V_{\rm s}$ is
the comoving survey volume, and $\mu$ is the cosine of the angle between
$\bm{k}$ and the line-of-sight direction.  Note that we have assumed
that the galaxy selection function is uniform over the redshift slice we
consider and ignored any boundary effects of survey geometry for
simplicity.

The first term in the bracket on the r.h.s of Eq.~(\ref{eqn:pkerror})
represents sampling variance.  Errors become independent of the number
density of galaxies when sampling variance dominates (i.e., $P_s\gg
\bar{n}_g$ over the range of $k$ considered), and thus the only way to
reduce the errors is to survey a larger volume.  On the other hand, the
second term represents shot noise, which comes from discreteness of
galaxy samples.  When shot noise dominates ($P_s\ll \bar{n}_g$), the
most effective way to reduce noise is to increase the number density of
galaxies by increasing exposure time per field.  Note that for a fixed
$\bar{n}_{g}$ the relative importance of shot noise contribution can be
suppressed by using galaxies with larger bias parameters, $b_1$, as
$P_s\propto b_1^2$.

We use the Fisher information matrix formalism to convert the errors on
$P_s(k)$ into error estimates of model parameters
\cite{Tegmark97,SE}. 
The Fisher matrix is computed from
\begin{eqnarray}
F_{\alpha\beta}&=&\frac{V_{\rm s}}{8\pi^2}\int^{1}_{-1}\!\!d\mu 
\int^{k_{\rm max}}_{k_{\rm min}}k^2dk~ 
\frac{\partial \ln P_s(k,\mu)}{\partial
p_\alpha}
\frac{\partial \ln P_s(k,\mu)}{\partial p_\beta}\nonumber\\
&&\times 
 \left[\frac{\bar{n}_gP_s(k,\mu)}{\bar{n}_gP_s(k,\mu)+1}\right]^2,
\label{eqn:fisher}
\end{eqnarray}
where $p_\alpha$ expresses a set of parameters. The partial derivative
 with respect to a given parameter $p_\alpha$ is evaluated around the
 fiducial model. 
 When combined with the
CMB constraints, we simply add the CMB and galaxy fisher matrices,
$F_{\mu\nu} =F^g_{\mu\nu}+F^{\rm CMB}_{\mu\nu}$, to obtain the joint
constraints.  The $1\sigma$ error on $p_\alpha$ marginalized over the
other parameters is given by
$\sigma^2(p_\alpha)=(\bm{F}^{-1})_{\alpha\alpha}$, where $\bm{F}^{-1}$
is the inverse of the Fisher matrix.  We follow the method described in
\cite{TJ} when we consider projected constraints in a two-parameter
subspace to see how the two parameters are correlated.

To calculate $F_{\alpha \beta}$ in Eq. (\ref{eqn:fisher}), we need to
specify $k_{\rm min}$ and $k_{\rm max}$ for a given galaxy survey. We
use the upper limit, $k_{\rm max}$, to exclude information in the
non-linear regime, where the linear theory prediction of density
fluctuations, Eq.~(\ref{eqn:ps}), becomes invalid.  Following \cite{SE},
we adopt the values of $k_{\rm max}$ for each redshift slices as given
in Table~\ref{tab:survey}.  As for the minimum wavenumber, we use
$k_{\rm min}=10^{-5}$ Mpc$^{-1}$, which gives well-converged results for
all the cases we consider.

\subsection{Model Parameters of a CDM Model}
\label{model}
The Fisher matrix formalism assesses how well given observables can
distinguish the true (``fiducial'') cosmological model from other
models.  The parameter forecasts we obtain depend on the fiducial model
and are also sensitive to the choice of free parameters. We include a
fairly broad range of the CDM dominated cosmology: the density
parameters are $\Omega_{\rm m}(=0.27)$, $\Omega_{\rm m}h^2(=0.14)$, and
$\Omega_{\rm b}h^2(=0.024)$ (note that we assume a flat universe); the
primordial power spectrum parameters are the spectral tilt, $n_s(=1)$,
the running index, $\alpha_s(=0)$, and the normalization parameter of
primordial curvature perturbation, $\delta_{\cal R}(=5.07\times
10^{-5})$ (the values in the parentheses denote the fiducial model).
The linear bias parameters, $b_1$, are included for each redshift slice
as given in Table~\ref{tab:survey}.

The dark energy parameters, the equation of parameter $w_0$ and the
sound speed $c_{\rm e}$, are allowed to vary in order to study how the
constraint on $c_{\rm e}$ changes with the assumed fiducial values.  We
also include the non-relativistic neutrinos as free parameters (see
\cite{TK} for the details). The neutrino oscillation experiments have
provided strong evidence that two or three of standard three-flavor
neutrinos have finite mass (e.g. see \cite{Vogel} for a review). We
assume throughout this paper that some of the three species are massive
and have become non-relativistic until the present.  The
non-relativistic neutrinos affect the structure formation by suppressing
the growth of matter density fluctuations at small spatial scales owing
to their large velocity dispersion. Hence, the non-relativistic
neutrinos induce a scale-dependent modification in the power spectrum
shape. For a range of the total neutrino mass implied from the neutrino
oscillation experiments, the effects of non-relativistic neutrinos and
dark energy clustering could be degenerate in a galaxy power spectrum,
especially when the dark energy sound horizon is comparable with the
neutrino free-streaming scale.  We shall below study this issue
quantitatively. The neutrino parameters we assume are the ratio of the
total neutrino mass density, $f_\nu(\equiv \Omega_{\nu}/\Omega_{\rm
m}=0.01)$, corresponding to the total neutrino mass $0.13$~eV, and the
number of non-relativistic neutrino species, $N_\nu^{\rm nr}(=2)$.  The
fiducial values of $f_\nu$ and $N_\nu^{\rm nr}$ are consistent with the
neutrino oscillation experiment results (see Fig.~3 in \cite{TK}) as
well as the upper bounds set from the cosmological experiments
\cite{Lahav02,Tegmark04,WMAP3}.  In summary, for a survey which consists
of $N_s$ redshift slices, we have $10+N_s$ parameters in total:
\begin{eqnarray}
p_\alpha&=&\left\{\Omega_{\rm m}, \Omega_{\rm m}h^2,\Omega_{\rm b}h^2, 
\delta_{\cal R}, n_s, \alpha_s, w_0, c_{\rm e},f_\nu,N_\nu^{\rm nr}\right.
\nonumber \\ 
&&\hspace{3em}\left. ,b_1(z_1),\cdots,b_1(z_{N_s})\right\}.
\end{eqnarray}

As shown in the literature (e.g., see \cite{WMAP3}), a galaxy survey
alone cannot determine all the cosmological parameters simultaneously,
but would leave some parameter combinations degenerated.  Combining the
galaxy survey constraints with the constraints from CMB temperature and
polarization anisotropies can be a powerful way to lift parameter
degeneracies \cite{EHT}.  When computing the Fisher matrix of CMB, we employ 7
parameters: 6 parameters (the parameters above minus the dark energy,
neutrino and bias parameters) plus the Thomson scattering optical depth
to the last scattering surface, $\tau(=0.10)$.  Note that our treatment
is rather conservative: we do not use any CMB constraints on the dark
energy parameters and the non-relativistic neutrinos.
The Fisher matrix for the joint experiment is
given by adding the CMB Fisher matrix to the galaxy Fisher matrix as
$F_{\alpha\beta}=F^{\rm g}_{\alpha\beta} +F^{\rm CMB}_{\alpha\beta}$.
We entirely ignore the contribution to the CMB from the primordial
gravitational waves.  We use the publicly-available CMBFAST code
\cite{cmbfast} to compute the angular power spectra of temperature
anisotropy, $C^{\rm TT}_l$, $E$-mode polarization, $C^{\rm EE}_l$, and
their cross correlation, $C^{\rm TE}_l$.  Specifically we consider the
noise per pixel and the angular resolution of the Planck experiment that
were assumed in \cite{EHT}.  Note that we use the CMB information in the
range of multipole $10\le l\le 2000$, and therefore we do not include
the ISW effect at low multipoles $l\simlt 10$ which the dark energy
perturbation may affect.

\section{Results}
\label{results}

\subsection{Forecasts for Testing Dark Energy Clustering}
\begin{figure}
  \begin{center}
    \leavevmode\epsfxsize=8.5cm \epsfbox{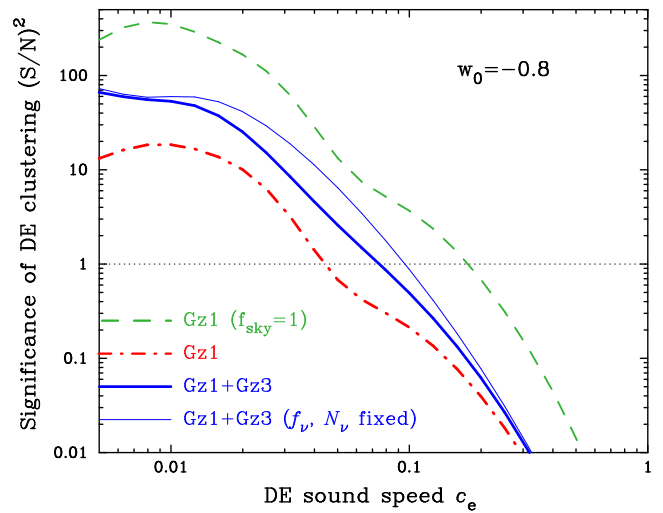}
  \end{center}
\vspace{-2em}
\caption{A significance of discriminating the dark energy clustering
 model from a smooth dark energy model ($c_e=1$) as a function of the
 fiducial value of $c_{\rm e}$, expected from the galaxy survey combined
 with Planck. 
 The fiducial value of the dark energy equation of state is kept fixed to
 $w_0=-0.8$. The bold solid curve shows the forecast for the nominal
 survey design of WFMOS (see Table~\ref{tab:survey}), while the
 dot-dashed curve shows the result when the Gz1 survey alone is used.
 The dashed curve shows the result for an ultimate all-sky survey of
 $z\sim 1$ galaxies.  \label{fig:ce} }
\end{figure}
As can be seen from Fig.~\ref{fig:pk}, the effect of dark energy
clustering on the linear power spectrum is only apparent for variations
that span an order of magnitude in $c_{\rm e}$ (also see
\cite{HuScranton} for the similar discussion). We therefore use
\begin{equation}
p_{c_{\rm e}}=\log_{10} c_{\rm e},
\label{eqn:pce}
\end{equation}
as a free parameter in the Fisher matrix calculation rather than
directly using $c_e$. Then, we define the following quantity to
estimate the significance of how well the future galaxy survey we
consider can discriminate between a given $c_{\rm e}$ model and a smooth
dark
energy  model with $c_{\rm e}=1$:
\begin{equation}
\left.\left(\frac{S}{N}\right)^2\right|_{c_{\rm e}}=
\left(\frac{\log_{\rm
 10}c_{\rm e}}{\sigma(\log_{10}c_{\rm e})}\right)^2,
\end{equation}
where $\sigma(\log_{\rm 10}c_{\rm e})\equiv
\sqrt{\left(\bm{F}^{-1}\right)_{p_{c_{\rm e}}p_{c_{\rm e}}}}$, the 68\%
C.L. error on $p_{c_{\rm e}}$ marginalized over the other parameter
uncertainties. 

The bold solid curve in Fig.~\ref{fig:ce} shows the forecasted
significance $(S/N)^2$ to discriminate between a dark energy clustering
model and the smooth model, for the WFMOS fiducial survey (Gz1 plus Gz3
in Table~\ref{tab:survey}) combined with Planck. Note that the fiducial
value of $w_0=-0.8$ is assumed here.  It is clearly seen that the dark
energy clustering can be distinguished at significance $(S/N)\simgt 1$
as the fiducial value of $c_{\rm e}$ gets smaller than $c_{\rm e}\approx
0.08$, where the dark energy sound horizon scale enters into a window
of the wavenumbers where the galaxy power spectrum can be precisely
measured with the galaxy survey (see  Fig.~\ref{fig:pk}).  
However, for too low sound speeds such as $c_{\rm e}\simlt 0.01$, $S/N$
saturates at $c_{\rm e}\simlt 0.01$ (or the dashed and dot-dashed curves
even have a peak around $c_{\rm e}\sim 0.01$). This is because the
transitions in $P_g(k,z)$ occur at $k$ comparable with or larger than
the maximum wavenumber $k_{\rm max}$ above which the linear power
spectrum information is not available due to the non-linearities in
structure formation.  On the other hand, the significance drops off
rapidly as $c_{\rm e}\rightarrow 1$ because the transitions occur at too
large spatial scales where the galaxy power spectrum measurement is
largely limited by the sample variance.  One may also notice that the
curves appear to have a jagged feature around $c_{\rm e}=0.05$. This
arises from degeneracies between dark energy clustering and
non-relativistic neutrinos.  This can be partly seen from the thin solid
curve, which shows the results when the neutrino parameters $f_\nu$ and
$N_\nu^{\rm nr}$ are fixed to the fiducial values; the jagged feature
disappears. We shall come back to this issue below in more detail.

The dot-dashed curve shows the result when the Gz1 survey alone is
combined with Planck.  Comparing the solid and dot-dashed curves
clarifies that the constraint on dark energy clustering is mainly from
the lower redshift survey compared to the $z\sim 3$ survey, because the
net dark energy density is more prominent at lower redshifts (see
Fig.~\ref{fig:pk}).  The dashed curve shows the forecast for an ultimate
full-sky redshift survey of $z\sim 1$, hypothetically obtained by
enlarging the survey area of $z\sim 1$ by a factor of 20. In this case,
the model with $c_{\rm e}\simlt 0.2$ can be detected at more than a
1-$\sigma$ level. This result might be compared with Fig.~8 in
\cite{HuScranton}; the authors investigated the ISW effect to probe dark
energy clustering via the CMB-galaxy cross-correlation measurements
assuming a full-sky CMB map and a full-sky, multi-color galaxy imaging
survey out to $z\sim 2$. Interestingly, our results show a similar-level
precision of constraining dark energy clustering to theirs, although our
method does not include the CMB information on dark energy clustering
(see Sec.~\ref{model}), and the survey area ($f_{\rm sky}\sim 0.05$) and
the redshift range are much smaller and narrower than theirs,
respectively.  This would be partly because a galaxy power spectrum
contains more independent modes of linear fluctuations in the
line-of-sight direction compared to the angular correlations. It is also
noticed that comparing the dashed curve with the dot-dashed (or solid)
curve shows about factor 20 increase in $(S/N)^2$, corresponding to the
gain in the sky coverage $f_{\rm sky}$. This implies that the constraint
on $c_{\rm e}$ is mainly from the galaxy survey, not as a result of
breaking the parameter degeneracies obtained by combining the galaxy
survey with CMB, as expected.  In summary, we conclude from the results
in Fig.~\ref{fig:ce} that a wide-field galaxy redshift survey by its own
offers an alternative, vital opportunity for testing the smoothness of
dark energy.

\begin{figure}
  \begin{center}
    \leavevmode\epsfxsize=8.5cm \epsfbox{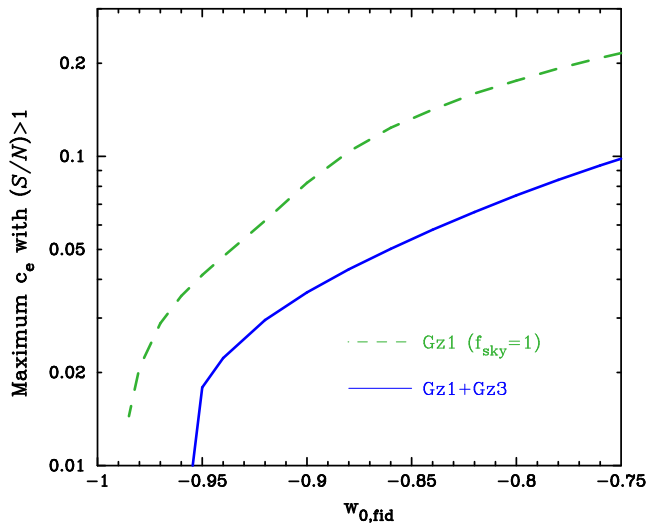}
  \end{center}
\vspace{-2em}
\caption{A maximum dark energy sound speed with $(S/N)|_{c_{\rm e}}\ge
1$ against the fiducial value of $w_0$. The solid and dashed curves are
the results for WFMOS (Gz1 plus Gz3) and the full-sky $z\sim 1$ survey,
respectively. As the underlying true cosmology approaches to the
cosmological constant model, one can detect dark energy clustering from
a galaxy survey only when the sound speed $c_{\rm e}$ is sufficiently
small.  \label{fig:ce-w} }
\end{figure}
Obviously, the ability to detect dark energy clustering from a galaxy
survey depends upon the net dark energy density or equivalently upon the
equation of state $w_0$ in our setting for a fixed $c_{\rm e}$, provided
$\Omega_{\rm de}$ and $\Omega_{\rm m}$ are tightly constrained.
Fig.~\ref{fig:ce-w} shows a maximum sound speed detectable with
$(S/N)\ge 1$ as a function of the given fiducial value of $w_0$. As dark
energy becomes close to the cosmological constant ($w_0=-1$), it is
getting more difficult to detect dark energy clustering: dark energy
clustering can be discriminated only if the sound speed is sufficiently
small so that the scale-dependent transitions in $P_g(k)$ are measurable
to within the measurement accuracies. To be more specific, the model
with $w_0\simgt -0.96$ allows a test of the dark energy smoothness for
the WFMOS survey, while the model with $w_0\simgt -0.98$ can be tested
for the ultimate full-sky survey.  Alternatively, if the true cosmology
has $w_0=-0.9$ ($-0.95$), the dark energy clustering can be detected
when $c_{\rm e}\simlt 0.02$ (0.04), while the full-sky survey allows the
detection when $c_{\rm e}\simlt 0.04$ (0.08), a factor of 2 improvement
over WFMOS.

\begin{table*}
\begin{center}
\begin{tabular}{l|ccccccccccccc}
\hline\hline
& $\Omega_{\rm m}$ & $\delta_{\cal R}$ & $w_0$ & $c_{\rm e}$
& $f_\nu$ & $N_{\nu}^{\rm nr}$ & $n_s$ &
$\alpha_{\rm s}$ & $\Omega_{\rm m}h^2$
& $\Omega_{\rm b}h^2$ & $\tau$ 
& $b_1(z=1)$& $b_1(z=3)$
\\ \hline
$c_{\rm e}$& 
$0.26$ 
& $-0.02$ 
& $0.18$ 
& 1 & $0.15$ & $-0.77$ & $-0.05$
 & $0.03$ & 0.23 & $-$0.23 & $-0.04$ & $-0.17$ & $-0.18$\\
\hline\hline
\end{tabular}
\end{center}
\caption{The correlation coefficients between the dark energy sound speed
parameter $p_{c_{\rm e}}$ (see Eq.~[\ref{eqn:pce}]) and the other parameter
$p_{\alpha}$ for the parameter error estimates for  WFMOS
combined with Planck.  The fiducial values for dark energy parameters
are set to $w_0=-0.8$ and $c_{\rm e}=0.05$.  The sound speed is most
degenerated with the non-relativistic neutrino parameter $N_\nu^{\rm nr}$.  }
\label{tab:coeff}
\end{table*}

\subsection{Parameter Degeneracies}

To more quantitatively understand the forecast for constraining the dark
energy clustering shown in the preceding subsection, it is useful to
examine how the parameters are degenerate with each other given the
survey data sets. Table~\ref{tab:coeff} gives the correlation
coefficients between the dark energy sound speed and the other parameters
for the WFMOS survey (Gz1 plus Gz3) combined with Planck. In the context
of the Fisher information matrix formalism, the correlation coefficient
is defined as
\begin{equation}
r\equiv \frac{(\bm{F}^{-1})_{p_{c_{\rm
e}}p_\alpha}}{\sqrt{({\bm{F}}^{-1})_{p_{c_{\rm e}}p_{{c_{\rm
e}}}}{(\bm{F}^{-1})_{p_\alpha p_\alpha}}}},
\end{equation}
where $p_\alpha$ denotes one of the model parameters.  When the
coefficient $|r|=1$, the two parameters are totally degenerate, while
$r=0$ means they are uncorrelated.  While the coefficients varies with
the fiducial model, we have assumed $w_0=-0.8$ and $c_{\rm e}=0.05$ for
the fiducial values of the dark energy parameters, where $c_{\rm
e}=0.05$ roughly corresponds to the dip scale of the curves in
Fig.~\ref{fig:ce} (also see below for a more detailed discussion).
Table~\ref{tab:coeff} shows that the dark energy sound speed is most
degenerate with the non-relativistic neutrino parameter $N_{\nu}^{\rm
nr}$ as expected, because a change in $N_{\nu}^{\rm nr}$ affects the
shape of linear power spectrum via the change in the neutrino
free-streaming scale\footnote{Note that the neutrino free-streaming
wavenumber is proportional to the absolute mass scale of
non-relativistic neutrino species and therefore depends on the neutrino
parameters of our interest as $k_{\nu, {\rm fs}}\propto f_\nu/
N_{\nu}^{\rm nr}$ (see \cite{TK}).}, but is insensitive to the $P(k)$
amplitude; the suppression rate in $P(k)$ amplitude due to the
neutrino free-streaming is primarily sensitive to another neutrino
parameter $f_\nu$ as $\Delta P(k)/P(k)\approx -8f_\nu$ in the
small-scale limit (see Fig.~2 in \cite{TK}).  On the other hand, the
sound speed is  only weakly correlated with the other
parameters. Encouragingly, therefore, while the parameters, $\Omega_{\rm
m}h^2, \Omega_{\rm b}h^2, n_s$ and $\alpha_s$, also modify the shape of
power spectrum, combining the galaxy survey with Planck is a very
powerful way to efficiently lift the parameter degeneracies (also see
\cite{TK} for the extensive discussion on the degeneracies between the
neutrino parameters and the primordial spectrum shape parameters).

It is also worth noting that the sound speed is only weakly correlated
with the parameters controlling the galaxy power spectrum amplitude such
as $\delta_{\cal R}$ and $b_1$, again verifying that the constraint on
$c_{\rm e}$ is primarily from the shape of galaxy poser spectrum rather
than the amplitude information.  Our method given in Sec.~\ref{ps}
includes the redshift distortion effect on the galaxy power spectrum,
which significantly helps break the degeneracy between the power
spectrum amplitude and the galaxy bias. We have checked that ignoring
the redshift distortion effect little changes our results.  Hence,
constraining $c_{\rm e}$ with the galaxy survey, if detected, is
generally robust against uncertainties involved in galaxy bias.

\begin{figure}
  \begin{center}
    \leavevmode\epsfxsize=8.5cm \epsfbox{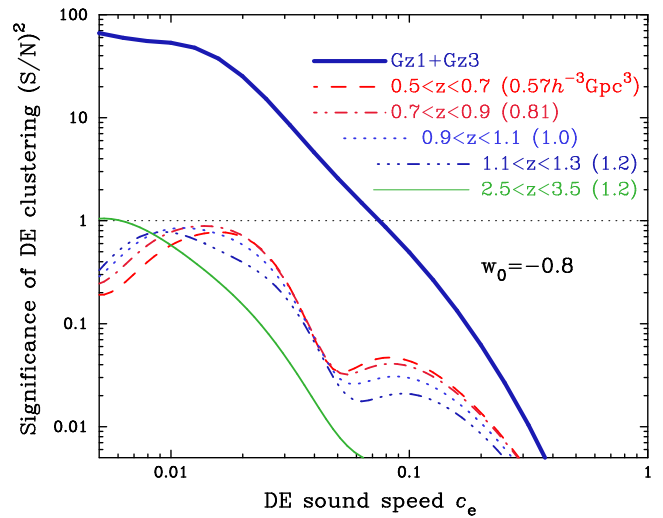}
  \end{center}
\vspace{-2em}
\caption{Shown is how each redshift slice of the WFMOS survey
 contributes to the dark energy constraint shown in
 Fig.~\ref{fig:ce}. The number in the parenthesis for the label of each
 curve denotes the comoving survey volume for each redshift slice (see
 Table~\ref{tab:survey}). The lowest redshift slice ($0.5\le z\le 0.7$)
 yields most contribution even though the slice has the smallest survey
 volume and the smallest $k_{\rm max}$, as dark energy is more prominent
 at lower redshifts. Also interesting is a jagged feature appears around
 $c_{\rm e}\approx 0.05$ when one redshift slice alone is used, while
 the jagged feature is significantly weakened when all the slices are
 combined. The reasons for this to happen are explained in text as well
 as the subsequent figures.  \label{fig:ce-z} }
\end{figure}

\subsection{Dark Energy Clustering and Massive Neutrinos}
\label{fnu-ce} 

In this subsection, we investigate in more detail the degeneracies
between the dark energy sound speed and the non-relativistic neutrinos.

In Fig.~\ref{fig:ce-z} we start with looking into the significance
$(S/N)^2$, as in Fig.~\ref{fig:ce}, but for the case when only one of
the redshift slices listed in Table~\ref{tab:survey} is used. All the
curves for each slice of Gz1 show a more prominent jagged feature around
$c_{\rm e}=0.05$ than the solid curve for all the slices combined,
implying that the constraint on $c_{\rm e}$ is weakened due to the
parameter degeneracies, because the dark energy sound horizon becomes
comparable with the neutrino free-streaming scale for a redshift range
accessible from the galaxy survey (we shall show this more explicitly
below).  However, most important result shown here is the parameter
degeneracies can be quite efficiently broken, when all the redshift
slices are combined, as shown by the bold solid curve. This implies that
a survey with wider redshift coverage can efficiently separate the two
effects from the measured $P(k)$.

\begin{figure}
  \begin{center}
    \leavevmode\epsfxsize=8.5cm \epsfbox{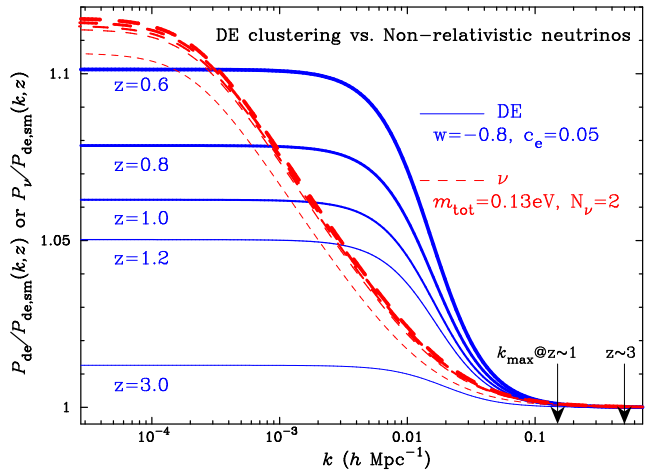}
  \end{center}
\vspace{-2em}
\caption{Effects of the dark energy clustering (solid curves) and the
 massive neutrinos (dashed) on the linear power spectrum shape, relative
 to the smooth dark energy model. The five curves from top to bottom are
 the results for $z=0.6, 0.8, 1.0, 1.2$ and $3$, respectively.  We
 assume $c_{\rm e}=0.05$, corresponding to the jagged feature in the
 curve of Fig.~\ref{fig:ce-z}, and $w_0=-0.8$ for the dark energy
 fiducial model and assume $f_\nu=0.01$ and $N_\nu^{\rm nr}=2$ for the
 neutrino mass and the number of non-relativistic neutrino species.
 Note that the dashed curves for the neutrinos are multiplied by an
 arbitrary factor so that $P_\nu(k)/P_{\rm de, sm}\rightarrow 1 $ at
 large $k$, taking into account the fact that the galaxy power spectrum
 amplitude involves uncertainty related to galaxy bias.  It is clear
 that the redshift dependences of the two effects are quite different.
 The two arrows in the horizontal axis denote $k_{\rm max}$ for the
 $z\sim 1$ and $z\sim 3$ surveys as given in Table~\ref{tab:survey}.
 \label{fig:pk_fnu-ce} }
\end{figure}
\begin{figure}
  \begin{center}
    \leavevmode\epsfxsize=8.5cm \epsfbox{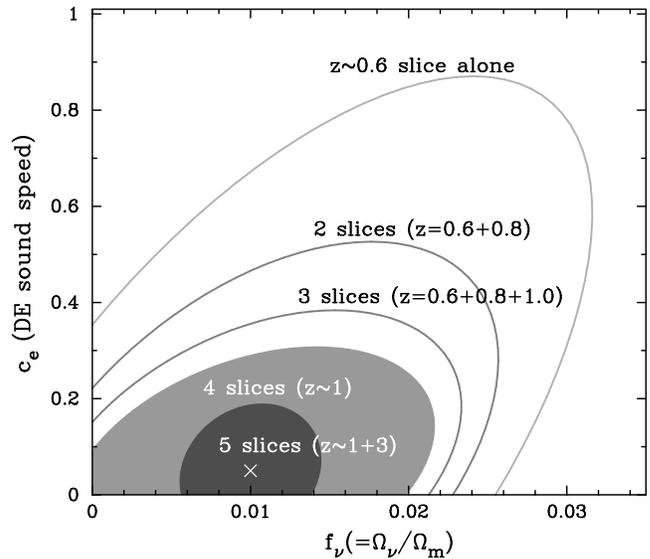}
  \end{center}
\vspace{-2em}
\caption{Projected $68\%$ ellipses in $f_\nu$ and $c_{\rm e}$
 sub-space. The error ellipses shrink more as more redshift slices are
 added. In particular, for the nominal WFMOS survey, a $z\sim 3$ survey
 helps break the degeneracy between the two parameters efficiently, even
 though the slice is insensitive to dark energy. Note that the prior
 $\sigma(w_0)=0.05$ is employed.  \label{fig:fnu-ce} }
\end{figure}

Figs.~\ref{fig:pk_fnu-ce} and \ref{fig:fnu-ce} give a more quantitative
explanation on the results in
Fig.~\ref{fig:ce-z}. Figure~\ref{fig:pk_fnu-ce} compares the effects of
the dark energy clustering and the non-relativistic neutrinos on the
linear power spectrum shape, for 5 redshift epochs. We have assumed
$w_0=-0.8$ and $c_{\rm e}=0.05$ for the fiducial values of dark energy
parameters as in Table~\ref{tab:coeff}.
It is intriguing to find the scale-dependent modifications in $P_g(k,z)$
due to these two effects are quite different: the dark energy clustering
induces sharper $k$-transitions in $P_g(k,z)$ as well as stronger
redshift dependence than the non-relativistic neutrinos do.  This
difference can be explained as follows. The neutrino free-streaming
effect on $P_g(k,z)$ has accumulated over a longer time duration since
the neutrino became non-relativistic, at $z_{\rm nr}\approx 240$ for
$m_{\nu,i}\approx 0.13$~ eV of our fiducial model, resulting in the
modification of the linear power spectrum shape over a wider range of
$k$. One the other hand, the dark energy clustering affects $P_g(k,z)$
only at low redshifts $z\simlt 1$.  One can thus utilize the different
redshift- and $k$-dependences to separate the two effect from the galaxy
power spectrum, if wider redshift and wavenumber coverages are
available.

Fig.~\ref{fig:fnu-ce} shows the projected $68\%$ C.L. contours in the
$f_\nu$-$c_{\rm e}$ sub-space, obtained when only one of the redshift
slices for WFMOS is used or some/all of the slices are combined. Note
that the prior $\sigma(w_0)=0.05$ was employed in this plot\footnote{We
have so far ignored the baryon acoustic peaks in $P(k)$, and including
the peaks is powerful to constrain the dark energy equation of state. To
take into account this, we assume the prior $\sigma(w_0)=0.05$ taken
from \cite{SE}. }. It is clear that, as a wider redshift coverage is
available, the degeneracy between $f_\nu $ and $c_{\rm e}$ determination
can be more broken. It is interesting to note that adding the $z\sim 3$
slice into the $z\sim 1$ slices can further shrink the error ellipses,
even though the dark energy clustering is negligible at $z\sim 3$. This
is because a higher redshift survey is more powerful for constraining
the neutrino parameters due to a gain in the linear fluctuations down to
larger $k$ (i.e. larger $k_{\rm max}$), allowing a more precise
measurement of the suppression in the small-scale power of $P(k)$ due to
the neutrino free-streaming.

\subsection{Effect of Dark Energy Clustering on $\sigma_8$}
\begin{figure}
  \begin{center}
    \leavevmode\epsfxsize=8.5cm \epsfbox{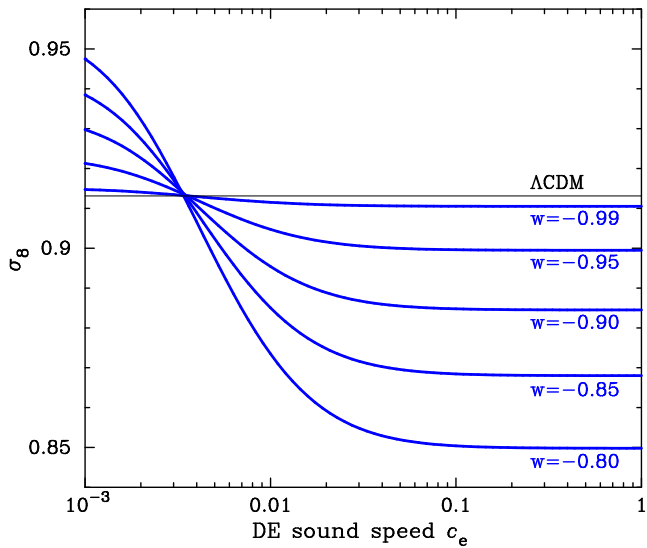}
  \end{center}
\vspace{-2em}
\caption{Difference in $\sigma_8$ when a generalized dark energy
 contribution is included and the power spectrum is normalized by the
 primordial curvature perturbation.  When $w_0>-1$ and the sound speed
 is large enough, $ \sigma_8$ is reduced compared to the $\Lambda$CDM
 prediction, because the structure formation slows down due to the more
 accelerated cosmic expansion at intermediate redshifts. On the other
 hand, when the sound speed is sufficiently small, the dark clustering
 enhances the power spectrum amplitude, yielding greater $\sigma_8$. The
 plot shows $c_{\rm e}\approx 0.005$ is the critical value of $c_{\rm
 e}$, where the two effects above cancel out so that the resulting
 $\sigma_8$ is independent of $c_{\rm e}$ and $w_0$, and similar to the
 $\Lambda$CDM prediction.  \label{fig:sigma8} }
\end{figure}
The cosmological parameter $\sigma_8$, the rms mass fluctuation today in
spheres of radius $8h^{-1}$~Mpc, is one of the most important parameters
for characterizing the clustering strength in large-scale structures at
low redshifts (e.g., \cite{bahcall}). However, for the moment $\sigma_8$
is relatively less accurately constrained compared to the other
parameters within the concordance CDM model. The three-year WMAP data
alone determines $\sigma_8$ to about $7\%$ accuracy \cite{WMAP3}. In
addition, methods based on independent data sets such as CMB, weak
lensing and Ly-$\alpha$ data sets show slightly inconsistent best-fit
values of $\sigma_8$, for example, as implied in Fig.~7 in \cite{WMAP3}
(also see \cite{Seljak06}): the WMAP favors slightly lower $\sigma_8$
than the methods looking at the local universe.

Since the CMB observables are primarily sensitive to the perturbations
at $z\sim 1000$, the present-day fluctuations, $\sigma_8$, are rather an
output parameter derivable from the parameters constrained by the CMB
spectra.  It is becoming common to use the primordial curvature
perturbation amplitude $\delta_{\cal R}$ to normalize the linear power
spectra for the CMB-based methods, as we have so far employed (see
Sec.~\ref{gps}).  Then, $\sigma_8$ can be computed once the linear power
spectrum shape and the growth rate from $z\sim 1000$ to present are
specified to within the measurement accuracies (e.g., see \cite{HuJain}
for a fitting formula of $\sigma_8$ for a flat CDM model with dark
energy of constant equation of state).  For this procedure, assuming a
model with dynamical dark energy could affect an estimate of $\sigma_8$
compared to the cosmological constant model because the generalized dark
energy could either suppress the growth rate due to the accelerated
expansion or add the dark energy perturbation to the power of total mass
perturbation. Which of these opposite two effects is dominant to
determine $\sigma_8$ depends on whether the dark energy sound horizon
scale is greater or smaller than $8h^{-1}$ Mpc.  Hence, if a discrepancy
between the $\sigma_8$ values inferred from the CMB and the local
universe observations is found, it may be as a result of the dark energy
contribution, and it would be worth to keep having attention to explore
such a signature from the cosmological probes at different redshifts.

Fig.~\ref{fig:sigma8} shows how the value of $\sigma_8$ changes as a
function of $c_{\rm e}$ and $w_0$, where other cosmological parameters
are kept fixed to the fiducial values.  When the sound speed $c_{\rm e}$
is large enough compared to $8h^{-1}$~Mpc, $\sigma_8$ is reduced with
increasing $w_0$ from $w_0=-1$ by slowing down the growth rate of mass
clustering. On the other hand, if the sound speed becomes sufficiently
small, albeit unrealistic, the dark energy clustering increases
$\sigma_8$ as a result of adding a power to the mass clustering at
scales relevant for $\sigma_8$.  Another interesting result is that the
critical sound speed $c_{\rm e, cr}\approx 0.005$ emerges where the
resulting $\sigma_8$ remains unchanged compared to the $\Lambda$CDM
prediction.

\section{Conclusion and Discussion}
\label{conc}

In this paper we have investigated the ability of a future galaxy
redshift survey to test the smoothness of dynamical dark energy, which
is another important consequence of a generalized dark energy with
$w(z)\ne-1$ and provides independent information on the nature of dark
energy from that carried by the equation of state. The dark energy
clustering signatures can be measured via a scale-dependent transition
in the power of the galaxy power spectrum appearing at scales comparable
with the dark energy sound horizon (see Fig.~\ref{fig:pk}).
It was shown that, for WFMOS survey, the sound speed can be detected
at more than a 1-$\sigma$ level, if the sound speed is small enough as
$c_{\rm e}\simlt 0.04 $ (0.02) when $w_0=-0.9$ ($-0.95$) (see
Figs.~\ref{fig:ce} and \ref{fig:ce-w}).  An effective way to improve the
ability is to enlarge the survey volume especially for low redshift
slices at $z\sim 1$. An ultimate full-sky survey could improve the lower
bound on the detectable $c_{\rm e}$ by a factor of 2.

Another interesting possibility of the future galaxy survey is the use
of the galaxy power spectrum to weigh the neutrino mass, as investigated
in \cite{TK}. We carefully investigate possible degeneracies between the
dark energy clustering and the non-relativistic neutrinos in the galaxy
power spectrum (see Fig.~\ref{fig:pk_fnu-ce} and also see
\cite{Hannestad} for the related discussion on the degeneracy between
the neutrino mass and the dark energy equation of state). We showed that
having a wider redshift coverage can efficiently separate the two
effects by utilizing the different redshift dependences; the dark energy
is prominent only at low redshifts $z\simlt 1$.  
In addition, modified gravity theories have received much recent
attention as an alternative explanation of cosmic acceleration without
dark energy, where gravity is weaker than the Einstein gravity at scales
comparable to the horizon (e.g., \cite{Shirata,Raul,Peacock,Fritz}). An
important test to discriminate this possibility from the dark energy
model is to simultaneously explore the cosmic expansion history and the
growth of structure formation from cosmological experiments, because the
two possibilities predict different growth rate even for an identical
expansion history. Once again, wider redshift coverage of a galaxy
redshift survey would be powerful to make a robust test of
discriminating between effects of dark energy clustering, massive
neutrinos, and modified gravity, which warrants a detailed study (also
see \cite{Peacock} for the discussion on the interplay between the
massive neutrinos and the modified gravity in the shape of mass power
spectrum).
We hope that the
results shown in this paper will give a useful guidance to designing a
future survey if the dark energy clustering is desired to pursue as one
of the science goals.

We have assumed throughout this paper that dynamical properties of the
dark energy are specified in terms of the equation of state and the
effective sound speed, with the adiabatic initial condition.  A merit of
this approach is we need not assume a specific form of the Lagrangian of
dark energy sector.  However, the limitation is there is no guarantee
this modeling can be applied to a full range of generic dark energy
models. For example, we have ignored a contribution from the trace-free
stress perturbation of dark energy, which is valid if the dark energy is
a scalar field (e.g. see \cite{Uzan} for the extension).  Since we have
little idea of what dark energy is, it is worth exploring a more general
modeling of dark energy properties and investigate the resulting effect
on structure formation.  Unless such a theoretical model is available,
we cannot extract all the cosmological information inherent in the
future high-precision cosmological data sets in an unbiased way.

Another non-trivial assumption we have made in this paper is the
adiabatic initial conditions for dark energy perturbations, without any
rigorous reasoning. If the dark energy field had existed since the
inflationary epoch in the early universe, the dark energy perturbations
naturally contained iso-curvature modes, and the effect just emerges at
low redshifts. Hence, a galaxy redshift survey could by its own open up
a new window for exploring the iso-curvature modes in structure
formation to probe the physics in the early universe, complementarily to
the CMB or the CMB-galaxy cross correlation
\cite{GordonHu,HuScranton}. Furthermore, if the primordial
non-Gaussianity (e.g. see \cite{Komatsu} for a thorough review) is
dominantly imprinted onto the sector of dark energy, the galaxy survey
allows us to explore the signal via, e.g., the bispectrum measurement of
galaxy clustering. These are our future study and will be presented
elsewhere.

The linear perturbation theory makes secure predictions on structure
formation, which allows an accurate interpretation of the cosmological
data sets from the detailed comparison. A main obstacle that a galaxy
redshift survey could contain in this procedure is uncertainties
involved in the galaxy biasing, where unknown, non-linear gastrophysics
may pollute the linear information even at large scales.  Although we
have assumed a scale-independent, linear bias, we do not think that a
more realistic bias such as a scale-dependent bias is a significant
contamination to the dark energy clustering constraint.  Because the
dark energy clustering induces characteristic scale-dependent
modification and strong redshift evolution for $P(k)$ (see
Fig.~\ref{fig:pk}), these properties can be very robust to discriminate
the dark energy clustering signals from the galaxy bias contamination or
more generally other systematics.
 
Finally, we comment on the non-linear gravitational clustering, which
could also provide a contamination to the linear theory
predictions. Encouragingly, however, the non-linear effects can be to
some extent corrected based on our refined knowledge of the structure
formation scenarios, based on the perturbation theory as well as the
N-body simulations \cite{Suto,Jain94,Jeong,Fritz}. Furthermore, as an
interesting possibility, coupling of different Fourier modes induced by
non-linear gravitational clustering could induce a transfer of the dark
energy clustering effect at large scales to the power of $P(k)$ at
smaller scales.  If this is the case to be apparent on the measured
power spectrum, an inclusion of the non-linearity in the model
predictions could improve a signal-to-noise to test smoothness of dark
energy, from the smaller-scale clustering information.
This issue will be worth exploring in detail, and will be presented
elsewhere.

\bigskip
{\it Acknowledgments:} The author thanks E.~Komatsu for valuable
discussions and also thanks for T.~Futamase, O.~Lahav, D.~Nitta,
P.~Norberg and Y.~Suto for helpful discussions. He also acknowledges the
warm hospitality of Caltech where this work was partly done.  This work
was supported in part by a Grand-in-Aid for Scientific Research
(17740129) of the Ministry of Education, Culture, Sports, Science and
Technology in Japan as well as the COE program at Tohoku University.  We
acknowledge the use of the publicly-available CMBFAST code
\cite{cmbfast}.

\end{document}